\documentclass[final,3p,times]{elsarticle}

\usepackage{amssymb}
\usepackage{amsmath}

\newcommand{\bfZ}{{\bf z}_{\perp}}

\newcommand{\bfk}{{\bf k}_{\perp}}

\newcommand{\Dp}{{\bf \Delta}_{\perp}}

\usepackage{xcolor}

\begin{document}
	
\begin{frontmatter}
		
		
		
\title{Chiral-odd generalized parton distributions of spin-1/2 baryons}
		

\author{Navpreet Kaur} 
\ead{knavpreet.hep@gmail.com}
		
\affiliation{organization={Department of Physics, Dr. B R Ambedkar National Institute of Technology Jalandhar},
		postcode={144008}, 
		 state={Punjab},
		country={India}}

\author{Monika Randhawa} 
\ead{monika@pu.ac.in}

\affiliation{organization={Department of Physics, Punjab University},
			city={Chandigarh},
	country={India}}
			
\author{Harleen Dahiya} 
\ead{dahiyah@nitj.ac.in}

\affiliation{organization={Department of Physics, Dr. B R Ambedkar National Institute of Technology Jalandhar},
	postcode={144008}, 
	state={Punjab},
	country={India}}
		
\begin{abstract}
	We present the tomographical structure of baryons by studying the nonforward matrix elements of lightlike correlation functions of the tensor current. At the leading twist, with the tensor current, four chiral-odd distributions are in count. We calculate these distributions in a diquark spectator model with light-front formalism by considering purely transverse momentum transfer, i.e., zero skewness. Predictions for the nucleons and light hyperons are studied, emphasizing the difference arising from their different quark flavors.
\end{abstract}
		
		
		
		
\end{frontmatter}

		
		
\section{Introduction}
\label{sec1}
		
An extensive information about the internal structure of hadrons can be obtained through the study of parton distribution functions. Among these, generalized parton distributions (GPDs) provide a unified framework of longitudinal momentum, transverse spatial distribution and spin structure of hadrons. In particular, the set of chiral-odd GPDs ($H_T$, $E_T$, $\tilde{H}_T$ and $\tilde{E}_T$) are sensitive to the transverse polarization of quarks inside hadrons and can be defined as the nonforward matrix elements of lightlike correlation functions of the tensor current \cite{Diehl:2001pm}. In the forward limit, at the leading twist, chiral-odd GPD $H_T$ reduces to the transversity distributions which has been studied experimentally and the results of their data analysis are predicted in Refs. \cite{Anselmino:2013vqa, Radici:2015mwa} for the case of proton. 
		
In the chiral-odd sector, a limited number of phenomenological studies have addressed the structure of light hyperons and at present, there is an experimental inaccessibility for their measurement \cite{Ledwig:2010tu, kucukarslan:2016xhx}.
Therefore, we present the study of chiral-odd GPDs of baryons. Particularly, focusing on the light hyperons (such as $\Sigma^+$ and $\Xi^0$) and their comparison with proton to emphasize the difference arising from their different quark flavors. 
For the present calculations, we have used the spectator diquark model based on the framework of light-cone dynamics, which is suitable for relativistic motion of hadrons. 
		
\section{Formalism}
		
We have considered asymmetric frame in which four-vector momentum coordinates of initial and final state of a baryon can be expressed as
\begin{eqnarray}
	P&=&\bigg(P^+,\frac{M^2}{P^+},\bf{0}_{\perp}\bigg), \nonumber \\
	P^\prime &=& \bigg((1-\xi)P^+,\frac{M^2+\Dp^2}{(1-\zeta)P^+},-\Dp \bigg) \, ,
\end{eqnarray}
respectively \cite{Goldstein:2013gra}. Here, $M$ represents the mass of a baryon and $\xi$ is the skewnwss parameter. The transverse momentum transfer is denoted by $\Delta_\perp$. In the spectator diquark model, a baryon is considered as a two body system, comprising of an active quark and a spectator diquark. The instant form SU(6) quark-diquark wave function with summation over different quark-diquark components can be defined as 
\begin{eqnarray} 
	|\mathcal{B}\rangle^{\lambda} = \cos \theta \sum_q a_q |q_1 ~\mathfrak{s}(q_2 q_3) \rangle ^{\lambda} + \sin \theta \sum_{q^\prime} b_q^\prime |q_1^\prime ~\mathfrak{a}(q_2^\prime q_3^\prime) \rangle ^{\lambda} \, .
	\label{InstantWfn}
\end{eqnarray}  
Fock state $|q_1 ~\mathfrak{s}(q_2 q_3) \rangle$ represents $q_1$ quark and $\mathfrak{s}(q_2 q_3)$ scalar diquark. Similarly, Fock state $|q_1^\prime ~\mathfrak{a}(q_2^\prime q_3^\prime) \rangle$ represents the $q_1^\prime$ quark and $\mathfrak{a}(q_2^\prime q_3^\prime)$ axial-vector diquark. The light-cone spin projections of the baryon is denoted by $\lambda$. The coefficients $a_q$ and $b_q^\prime$ satisfy the normalization condition $\cos^2 \theta \sum_q a_q+\sin^2 \theta \sum_{q^\prime} b_q^\prime=1$ \cite{Lichtenberg:1968zz}. For nucleons and light hyperons, only the associated coefficients have different values for different quark-diquark pair. The mixing angle $\theta$ arises due to the spin-flavor SU(6) symmetry breaking. The light-cone wave functions for scalar and axial-vector diquark is defined by
\begin{eqnarray}
	\psi^{\lambda}_{\lambda_q} (x,\bfk) = \sqrt{\frac{k^+}{(P-k)^+}} \frac{1}{k^2-m^2_q} \bar{u} (k,\lambda_q) ~\mathcal{Y}_{\mathfrak{s}}~ U(P,\lambda),
	\label{ScalarWfn}
\end{eqnarray}
\begin{eqnarray}
	\psi^{\lambda}_{\lambda_q \lambda_{\mathfrak{a}}}  (x,\bfk) = \sqrt{\frac{k^+}{(P-k)^+}} \frac{1}{k^2-m^2_q} \bar{u} (k,\lambda_q) \epsilon^\ast_\mu (P-k,\lambda_{\mathfrak{a}}) \cdot \mathcal{Y}_{\mathfrak{a}}^\mu~ U(P,\lambda) ,
	\label{VectorWfn}
\end{eqnarray}
respectively \cite{Bacchetta:2008af}. Here, the terms $~\mathcal{Y}_{\mathfrak{s}}$ and $\mathcal{Y}_{\mathfrak{a}}^\mu$ correspond to the scalar and axial-vector vertex functions with $\epsilon^\ast_\mu (P-k,\lambda_{\mathfrak{a}})$ as four-vector polarization of spin-$1$ diquark, carrying momentum ($P-k$) and helicity $\lambda_a$. The terms $m_q$, $\lambda_q$ and $k$ represent the mass, helicity and momentum carried by an active quark, respectively. The quark-quark correlator with tensor current is defined as
\begin{eqnarray}
	F_{\lambda \lambda^\prime}=\frac{1}{2} \int \frac{dz^-}{2\pi} e^{i \bar{x}\bar{P}^+z^-} \bigg\langle P^\prime,\lambda^\prime \bigg| \bar{\psi}\bigg(-\frac{z}{2}\bigg) \sigma^{+j} \gamma_5 \psi \bigg(\frac{z}{2} \bigg) \bigg|P,\lambda \bigg\rangle \bigg|_{z^+=0, \bfZ=0},
	\label{CO_correlator}
\end{eqnarray}
which is parameterized by
\begin{eqnarray}
	F_{\lambda \lambda^\prime} &=& \frac{1}{2P^+} \bar{U}(P^\prime,\lambda^\prime) \bigg[H_T(x,\xi,t) \sigma^{+i} \gamma_5 + \tilde{H}_T(x,\xi,t) \frac{\epsilon^{+j \alpha \beta} \Delta_\alpha \bar{P}_{\beta}}{M^2} + E_T(x,\xi,t) \frac{\epsilon^{+j \alpha \beta} \Delta_\alpha \gamma_\beta}{2M} \nonumber \\&+&  \tilde{E}_T(x,\xi,t) \frac{\epsilon^{+j \alpha \beta} \bar{P}_{\alpha} \gamma_\beta}{M} \bigg] U(P,\lambda).
	\label{Corr1}
\end{eqnarray}
Here, transverse index is denoted by $j(=1,2)$ \cite{Diehl:2001pm}. The quantities $\bar{U}(P^\prime,\lambda^\prime)$ and $U(P,\lambda)$ denoted the final and initial state spinor of a baryon, respectively. $\bar P$ corresponds to the baryon's average momentum and the invariant momentum transfer is denoted by the term $t$ (GeV$^2$). 
\section{Results and Discussions}
\begin{figure}[t]
\centering
\begin{minipage}[c]{0.98\textwidth}
	(a)\includegraphics[width=4.75cm]{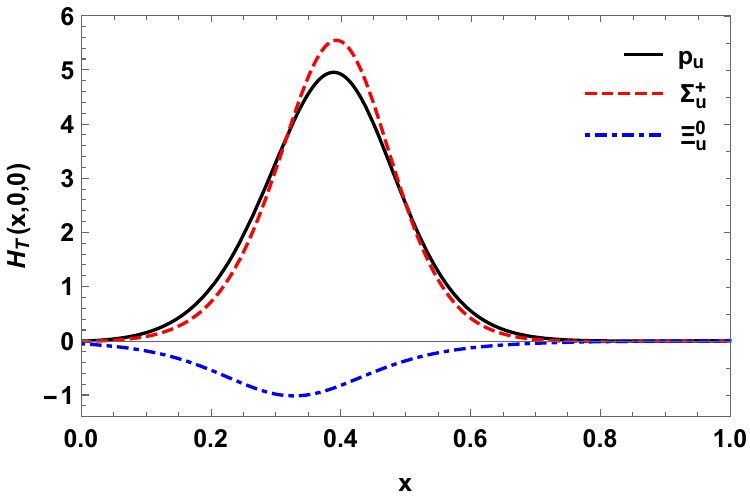}
	\hspace{0.03cm}
	(b)\includegraphics[width=4.75cm]{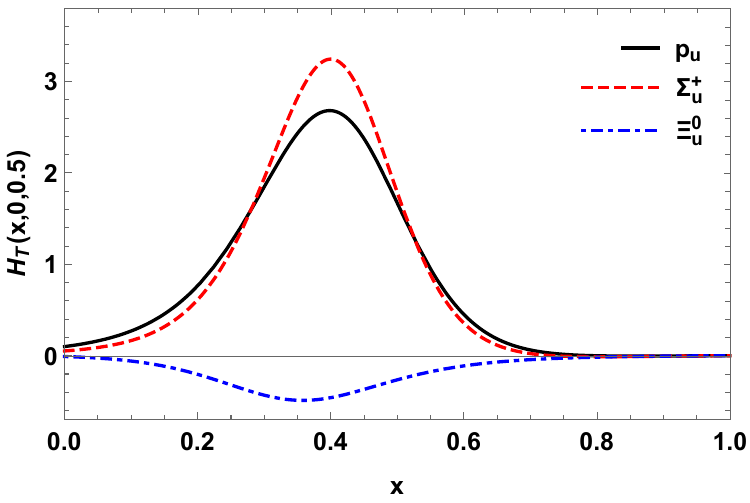}
	\hspace{0.03cm}	 
	(c)\includegraphics[width=4.75cm]{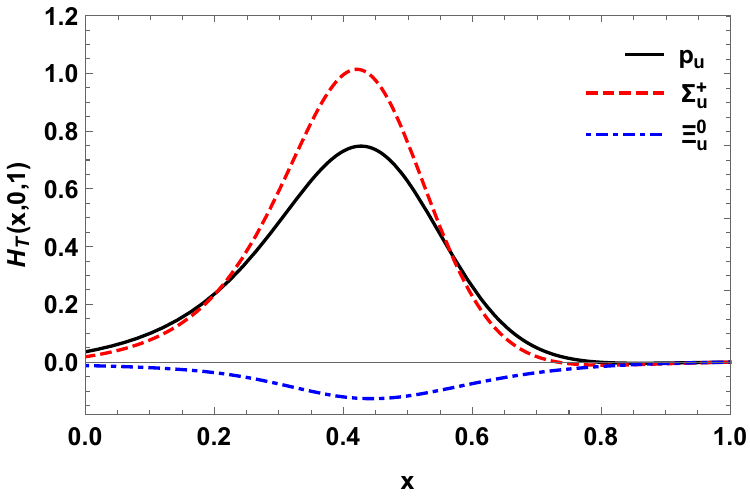}
	\hspace{0.03cm}
\end{minipage}
\caption{\label{fig1} Comparison of chiral odd GPDs $H_T(x,\xi,t)$ as a function of longitudinal momentum fraction $x$ at fixed transverse momentum transfer (a) $-t=0$ (GeV$^2$), (b) $-t=0.5$ (GeV$^2$) and (c) $-t=1$ (GeV$^2$) for $u$ quark flavor of proton, $\Sigma^+$, and $\Xi^0$.}
\end{figure}

The results of chiral-odd GPDs for zero skewness ($\xi=0$) have been obtained by using the numerical parameters mentioned in Ref. \cite{Kaur}. At $\xi=0$, the GPD $\tilde{E}_T(x,\xi,t)$ gives zero as it shows odd behavior under skewness as a consequence of time-reversal invariance and $-t=\Delta_\perp^2$. Comparison chiral-odd GPD $H_T(x,,0,0)$ of valence $u$ quark flavor as a function of longitudinal momentum fraction $x$ for zero momentum transfer throughout the process among proton and light hyperons ($\Sigma^+$ and $\Xi^0$) is shown in Fig. \ref{fig1}(a), which correspond to their transversity distributions. The contrast behavior is shown by the $\Xi^0$ hyperon as its peak value lies on the smaller of $x$ and has a smaller magnitude than the proton and $\Sigma^+$. The overall negative distribution indicates that for a transversely polarized $\Xi^0$, the measured $u$ quark is more likely to be found with anti-aligned spin with respect to $\Xi^0$. Between proton and $\Sigma^+$, proton is found to have a smaller magnitude, but the peak value lies very close to each other, which is attributed to the compensation of the increased mass of the baryon with its spectator diquark. Further, on increasing the value of transverse momentum transfer $t$ (GeV$^2$), the distribution of each baryon follows a generic trend of falling magnitude with a shift in peak value to higher values of $x$ as shown in Figs. \ref{fig1}(b) and \ref{fig1}(c). However, the change is more rapid for the case of protons than light hyperons as the value of $t$ (GeV$^2$) increases. 

\begin{figure}[t]
	\centering
	\begin{minipage}[c]{0.98\textwidth}
		(a)\includegraphics[width=4.75cm]{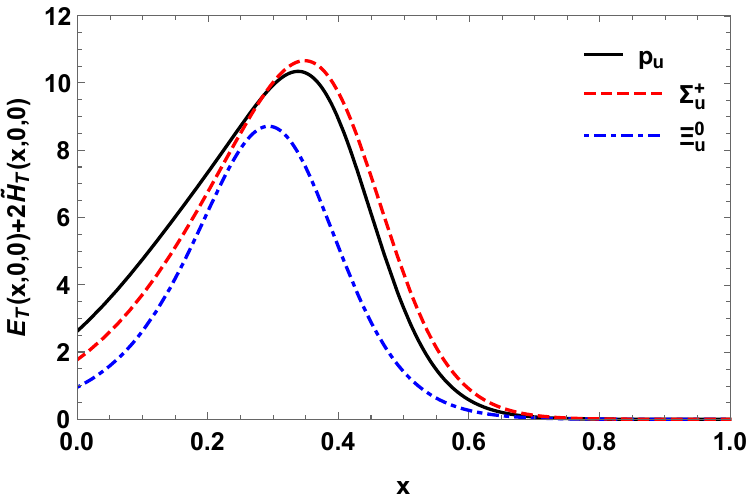}
		\hspace{0.03cm}
		(b)\includegraphics[width=4.75cm]{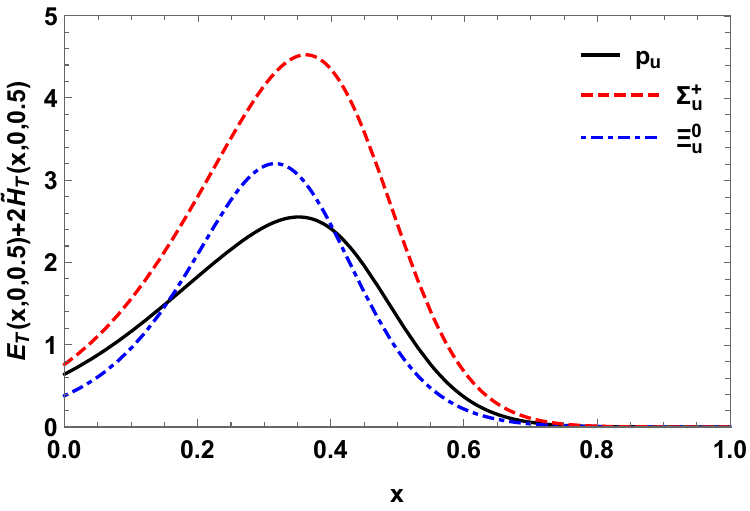}
		\hspace{0.03cm}	 
		(c)\includegraphics[width=4.75cm]{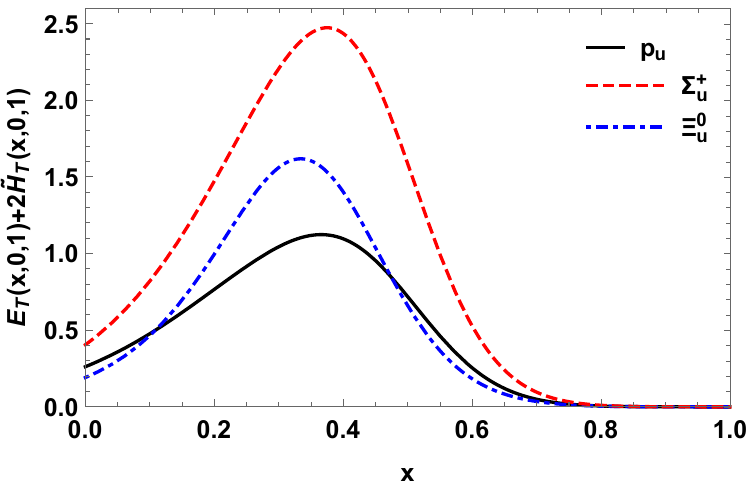}
		\hspace{0.03cm}
	\end{minipage}
	\caption{\label{fig2} Comparison of chiral odd GPDs $E_T(x,\xi,t)+2 \tilde{H}_T(x,\xi,t)$ as a function of longitudinal momentum fraction $x$ at fixed transverse momentum transfer (a) $-t=0$ (GeV$^2$), (b) $-t=0.5$ (GeV$^2$) and (c) $-t=1$ (GeV$^2$) for $u$ quark flavor of proton, $\Sigma^+$, and $\Xi^0$.}
\end{figure}

Comparison of valence $u$ quark flavor for the combination of chiral-odd GPDs $E_T(x,,0,0)+2\tilde{H}(x,0,0)$ as a function of longitudinal momentum fraction $x$ for zero momentum transfer throughout the process among proton and light hyperons is shown in Fig. \ref{fig2}(a). At $x=0$, these distributions have finite values, which is consistent with the results of Ref. \cite{Kaur:2023lun} for the case of the proton. Among proton, $\Sigma^+$ and $\Xi^0$, again the hyperon $\Xi^0$ is found to have a smaller magnitude and peaks at a smaller value of $x$. The effect of the increment of $t$ (GeV$^2$) can be seen among the subplots of Fig. \ref{fig2} as the distributions corresponding to $t=0.5$ (GeV$^2$) and $t=1$ (GeV$^2$) are represented in Figs. \ref{fig2}(b) and \ref{fig2}(c). The behavior of the decrement of magnitude and shifting of peak on higher value of $x$ is found to be the same among proton, $\Sigma^+$ and $\Xi^0$, with faster declining rate for proton as observed for the case of chiral-odd GPD $H_T(x,0,t)$.
\section*{Acknowledgments}
H.D. would like to thank  the Science and Engineering Research Board, Anusandhan-National Research Foundation, Government of India under the scheme SERB-POWER Fellowship (Ref No. SPF/2023/000116) for financial support and International Travel Support (File number: ITS/2025/005034).

	\end{document}